\documentclass[journal]{IEEEtran}
\usepackage{cite}
\usepackage{graphicx}
\usepackage{subfigure}
\usepackage{amsmath}
\usepackage{amsfonts,amssymb}
\usepackage{algorithm}
\usepackage{algorithmic}

\usepackage{array}
\usepackage{multirow}
\usepackage{enumerate}
\usepackage{color}

\makeatletter
\newcommand{\removelatexerror}{\let\@latex@error\@gobble}
\makeatother

\def\BibTeX{{\rm B\kern-.05em{\sc i\kern-.025em b}\kern-.08em
		T\kern-.1667em\lower.7ex\hbox{E}\kern-.125emX}}

\begin{document}
\title{Scenario-Agnostic Deep-Learning-Based Localization with Contrastive Self-Supervised Pre-training }

\author{
Lingyan Zhang,~\IEEEmembership{Member,~IEEE},~Yuanfeng~Qiu, Dachuan Li,~Tingting Zhang,~\IEEEmembership{Member,~IEEE,}\\Shaohua Wu,~\IEEEmembership{Member,~IEEE,} and Qinyu Zhang,~\IEEEmembership{Senior Member,~IEEE}
        




}


\maketitle

\begin{abstract}
Wireless localization has become a promising technology for offering intelligent location-based services. Although its localization accuracy is improved under specific scenarios, the short of environmental dynamic vulnerability still hinders this approach from being fully practical applications. In this paper, we propose CSSLoc, a novel framework on contrastive self-supervised pre-training to learn generic representations for accurate localization in various scenarios. Without the location information supervision, CSSLoc attempts to learn an insightful metric on the similarity discrimination of radio data, in such a scenario-agnostic manner that the similar samples are closely clustered together and different samples are separated in the representation space. Furthermore, the trained feature encoder can be directly transferred for downstream localization tasks, and the location predictor is trained to estimate accurate locations with the robustness of environmental dynamics. With extensive experimental results, CSSLoc can outperform classical and state-of-the-art DNN-based localization schemes in typical indoor scenarios, pushing deep-learning-based localization from specificity to generality. 
\end{abstract}


\section{Introduction}
With the requirement of inherently intelligent capability for 6G networks \cite{6GPos}, deep-learning-based wireless localization has become a key-enabled technology for offering high-quality location-based services (LBS)  and location-aware applications \cite{AutoLoc,expr2022,AIWiFi}, such as  localization and tracking \cite{Yassin2017Recent,gomap,DeFi}, autonomous vehicles \cite{AutoLoc}, internet of vehicles (IoV) \cite{Siamese}, 
  and so on. Compared with other approaches based on distance and angle estimation \cite{Yassin2017Recent,DeFi}, related solutions adopt a data-driven manner to own the superiority of easy radio availability, simple system implementation, and the appealing tradeoff between system overhead and localization performance \cite{expr2022,FLSurvey,AIWiFi}. 
Due to the strong representation capability of deep neural networks (DNNs), current schemes \cite{DeepFi,WiDeep,CNNLoc,RNN, Swin-Loc}   can extract finer features to characterize the relationships between the radio features and users' locations in complicated multipath environments. 
Then, the location estimation is determined by optimal feature matching in the existing radio map which is constructed by the extensive collection of radio signals with the corresponding location information, i.e., radio fingerprints \cite{FLSurvey,AIWiFi}. 


While promising localization accuracy has been achieved under specific scenarios, it is still challenging for the deep-learning-based approach to acquire environmental adaptability and generalization ability for practically large-scale applications  \cite{l2l,FLSurvey,AuMc,IPSSurvey}.
Practically, radio signals are highly vulnerable to environmental dynamics that involve user motion, transient interference, and surrounding changes in temperature, humidity, and weather conditions \cite{IPSSurvey,AuMc,l2l}. The temporal fluctuations of received signal strength (RSS) are further exaggerated with dense multipath propagation, such as reflection, diffraction, scattering, and shadowing in complex environments, and further, the feature distribution is gradually out of date, even broken down over time \cite{AIWiFi,l2l,FLSurvey}. 
Hence, the query features captured by real-time radio measurements are disparate from the extracted features in the existing radio map, grossly deteriorating the accuracy of location estimations. 
Due to the uncertainty of RSS with unpredictable environmental dynamics, the intuitive solution repeats the site survey process to periodically update the radio map with cumbersome maintenance costs in both labor and time \cite{AuMc,FLSurvey}. 


Advanced works \cite{l2l,Siamese,MatchNet,MetaLoc,DeepWiPos,GraphLoc,RobLoc,DadLoc,DRF} realize an algorithm-level adaptation to learn environment-invariant representations by exploring and exploiting the robust factors underlying radio fingerprints with complicated environment dynamics. 
Since radio variances from the neighboring fingerprints remain relatively stable, even though RSS incur uncertain fluctuations with environmental dynamics, 
CSI-SCNN \cite{Siamese} and LESS \cite{l2l} design Siamese neural networks to learn the metric of feature similarity for modeling local proximity property among different radio fingerprints. 
GraphLoc \cite{GraphLoc} encodes the structural information underlying radio fingerprints to learn neighborhood patterns by developing graph neural networks (GNN) with an attention-based mechanism for robust indoor localization. 
Furthermore, deep-transfer-learning-based schemes \cite{RobLoc,DadLoc,DRF,ILOT} are proposed to learn transferable representations by minimizing the distribution discrepancy caused by environmental dynamics. 
Although these localization solutions can acquire environmental adaptability with a reasonable localization performance, the feature representations are learned relying on the extensive collection of radio fingerprints with the burdensome effort of an initial system deployment as well. 

To address these issues, we propose a novel framework of deep-learning-based localization with \textbf{c}ontrastive \textbf{s}elf-\textbf{s}upervised learning, namely CSSLoc, that can learn generic representations for downstream localization tasks at lower deployment and maintenance costs in various indoor environments. 
Instead of the feature extraction with the location information supervision, CSSLoc focuses on 
learning the similarity discrimination in a so-called scenario-agnostic way that the similar samples are closely clustered together and the different samples are far away \cite{CRLreview}.  
We revisit supervised-learning-based localization process and find out that the learned future representations can automatically capture the characteristics of the apparent similarity and dissimilarity underlying radio data themselves, not from the location annotations or the location information \cite{InstDisc,InstSpread}. It inspires us to learn an insightful metric on the similarity of radio samples in the representation space. 
On this basis, we design Siamese neural networks to train unannotated radio data and attempt to learn to discriminate their inherent similarities and dissimilarities. The learned representations can be directly transferred to various downstream tasks, such as not only indoor localization in our case but also wireless sensing as activity recognition, device-free localization, etc.


Specifically, we first build the contrastive self-supervised pre-training model with centralized radio data in the cloud to characterize their inherent similarities and dissimilarities in the latent space. 
By using multi-antenna and multi-subcarrier radio measurements, we leverage CSI samples to generate radio images with the pixel level of temporal and spatial correlation \cite{DadLoc,DRF,ILOT}. 
The feature vectors are obtained by the feature encoder with convolutional neural networks (CNN), and the learnable nonlinear transformation is performed to improve the quality of the feature representations in the low-dimensional embedding space. 
Depending on the similarity discrimination, the contrastive loss is formulated to 
enable the encoded query to match with its positive embeddings, which clusters the similar embeddings together and pushes the dissimilar ones away. 
During contrastive self-supervised pre-training, CSSLoc adopts a moving-averaged update to optimize the feature encoder, which can keep the similarity metric between the feature embeddings as consistent as possible. 
Furthermore, for downstream localization tasks, the learned feature encoder is directly transferred to capture the appropriate features, and the location predictor is trained with limited  CSI fingerprints in an indoor scenario, effectively reducing the deployment efforts.  
When real-time CSI measurements are received to generate radio images in a specific scenario, the target location is estimated by the softmax outputs using a probabilistic method. 
With extensive experimental results, CSSLoc can outperform both classical and state-of-the-art DNN-based localization schemes with better system performance at lower deployment and maintenance costs.  

The main contributions of the proposed CSSLoc system can be summarized as follows.
\begin{enumerate}
	\item We propose a novel framework of deep-learning-based localization with contrastive self-supervised learning that can learn generic representations to characterize the similarity and dissimilarity of radio data in the representation space, facilitating DNN-based localization technology from specificity to generality.

	\item We provide a fresh perspective to learn an insightful metric on the similarity discrimination in an environment/scenario-agnostic manner. Compared with the existing approaches, CSSLoc can narrow the gap between system overhead and cross-scenario localization with high performance.  
	
	\item We validate the effectiveness of the CSSLoc prototype in typical indoor environments with extensive performance evaluations. Satisfactory localization performances can make great progress in deep-learning-based localization technology to provide fully practical LBS.
\end{enumerate}

The rest of this paper is structured as follows. Related work are reviewed in Section II. We present our motivation and key idea in Section III. 
The system design and implementation of the proposed CSSLoc scheme are elaborated in Section IV. In the next section, the effectiveness of the developed CSSLoc prototype is validated with an extensive experimental study. Finally,  we draw the conclusion in Section VI.

\section{Related Work}
Deep-learning-based localization has become a promising technology of providing intelligent LBS for ubiquitous applications, which has attracted intensive research interests \cite{FLSurvey,IPSSurvey,Yassin2017Recent,expr2022}. Early works design classical DNN, such as denoising auto-encoder \cite{WiDeep}, CNN \cite{CNNLoc}, and recent neural networks \cite{RNN}, to learn deep representations which build the mappings between the signal features and location coordinates. Swin-Loc \cite{Swin-Loc} designs a Transformer-based network to learn temporal-spatial representations to enhance localization accuracy.
Based on supervised learning, they obtain the environment-dependent features without the robustness of environmental dynamics, which leads to a shortage of system generalization for offering practical LBS. 
On par with feature learning, DeepFi \cite{DeepFi} adopts an unsupervised-learning framework to learn feature representations by reconstructing raw radio signals. WiGAN \cite{WiGAN} employs generative adversarial networks to learn feature representations using a maxi-min game. 
However, they are still relevant to the interested environment with unpredictable wireless propagation and complex surrounding changes. 
For accurate location estimations, it is inevitable to periodically conduct site surveys for radio map updates at high maintenance costs \cite{AIWiFi,FLSurvey,IPSSurvey}.  

Recently, the primary concern for deep-learning-based localization technology is to acquire environmental adaptability in complicated changing environments/scnearios \cite{AIWiFi,IPSSurvey,FLSurvey}. Mainstream schemes aim to learn environment-invariant representations by refining the robust factors underlying radio fingerprints with environmental dynamics \cite{FLSurvey}. 
Adaptive localization schemes  \cite{Siamese,l2l} disclose that  radio variances among the neighboring fingerprints remain relatively stable, even though RSS incurs uncertain fluctuations with the environmental changes, and thus, Siamese neural networks are designed to learn relation representations with the similarity metrics of the radio features. Meanwhile, GNN is harnessed to encode the structural information of radio fingerprints for learning neighborhood patterns \cite{GraFin}, and further robust localization schemes \cite{GNN-Loc, GraphLoc} develop graph attention networks with the residual structure to attain an accuracy improvement in indoor environments. Alternatively, RobLoc \cite{RobLoc} incorporates the empirical maximum mean discrepancy (MMD) distance into DNN for learning transferable representations to eliminate the distribution discrepancy caused by environmental dynamics. 
Pioneer works \cite{DRF, ILOT} adopt domain adversarial adaptation networks to learn domain-invariant representations by maximizing the confusion of feature distributions across different environments (domains). 
DadLoc \cite{DadLoc} develops  dynamic adversarial adaptation networks to learn the evolving feature representations with the environmental dynamics. Although these localization solutions can acquire the environmental adaptability to achieve a reasonable localization performance, robust feature representations are still learned by the initial construction of a radio map with burdensome deployment efforts. 

Therefore, we propose a novel framework of deep-learning-based localization with contrastive self-supervised learning to push this approach from specificity to generality at lower deployment and maintenance costs. Instead of location information supervision, 
CSSLoc endeavors to learn meaningful representations which can cluster the similar embeddings together and separate the dissimilar ones away in the scenario-agnostic way. The learned representations are directly transferred for downstream localization tasks in various interested environments.

\section{Problem Formulation and Motivation}
Generally, deep learning-based localization \cite{DeepFi, WiDeep, CNNLoc} can be considered as a classification problem that involves offline training and online localization. In the offline training stage, radio fingerprints are collected to extract appropriate features with a site survey.
The physical area of the interested scenario is gridded to deploy the reference points (RPs) and record their location coordinates $\{L_1,  L_2, \ldots, L_R\}$ where $R$ is the number of RPs. The radio signals are measured in every RP and the radio map is constructed as $\mathcal{D}=\{\mathbf{x}_i,y_i\}_{i=1}^N$ where $\mathbf{x}_i$ is the CSI amplitude vector in our case with its location annotation $y_i$ (its location coordinate $L_{y}$), and $N$ is the total number of the training CSI samples. 
Then, DNN with the learnable parameters $\theta$ are designed to learn deep representations $f(\mathbf{x})$ and the location predictor $p$ is trained with the softmax classification in a specific environment as
\begin{equation}	
	\boldsymbol{\theta}^{*}=\arg\min_{\boldsymbol{\theta}}\frac{1}{N}\sum_{i=1}^{N}\ell_{ce}\left(p(f(\mathbf{x}_i)), y_i \right),
\end{equation} 
where $\ell_{ce}$ is the cross-entropy loss function. 
By the location information supervision, however, the learned environment-dependent representations hinder this approach from providing follow-up LBS with the time-consuming system deployment in diverse scenarios \cite{FLSurvey,AIWiFi}. 

\begin{figure}[tbp]
	\centering
	\subfigure[]{
		\includegraphics[width=0.11\textwidth]{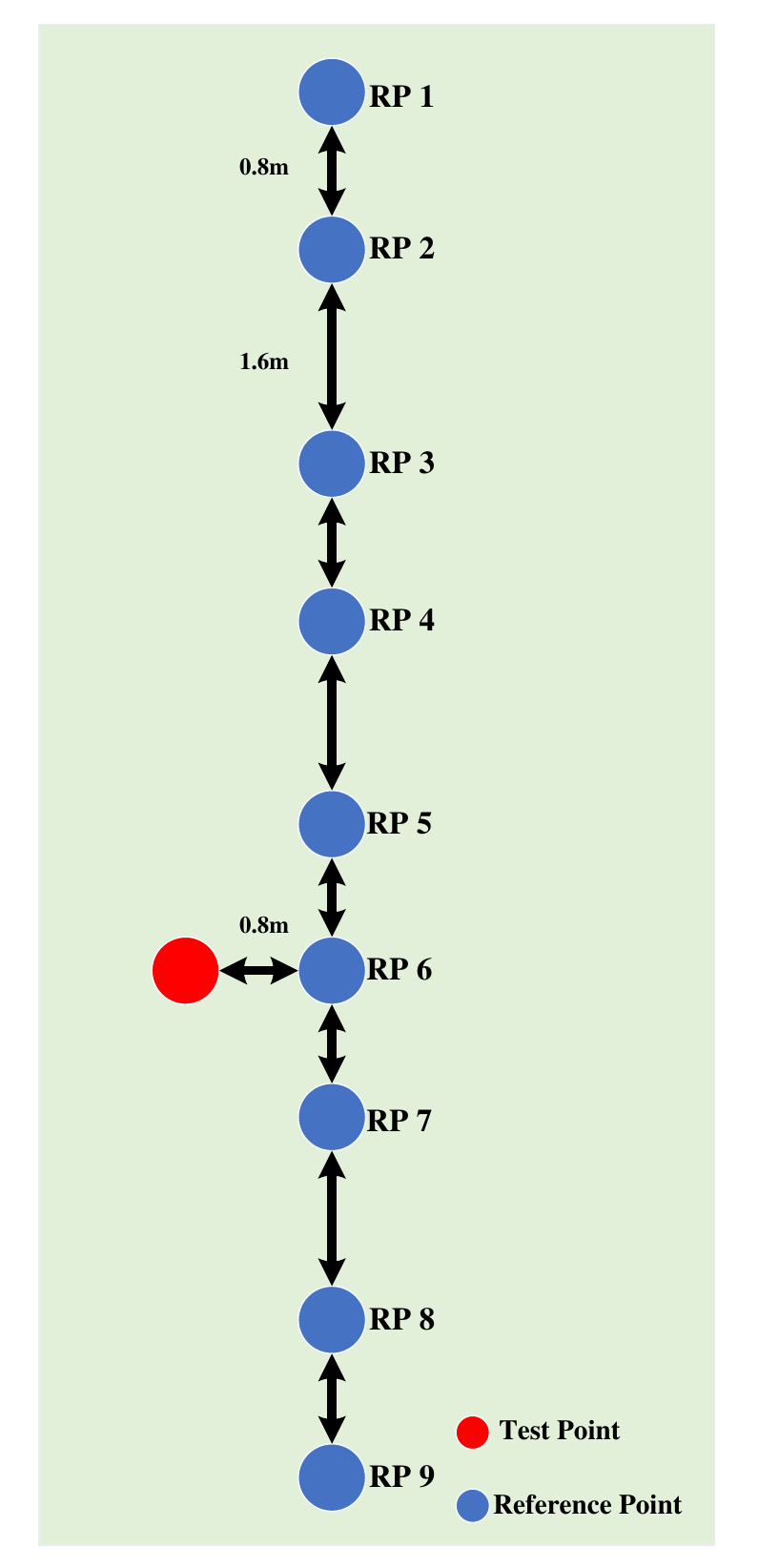}}
	\subfigure[]{
		\includegraphics[width=0.3\textwidth]{distributions.pdf}}
	\subfigure[]{
		\includegraphics[width=0.45\textwidth]{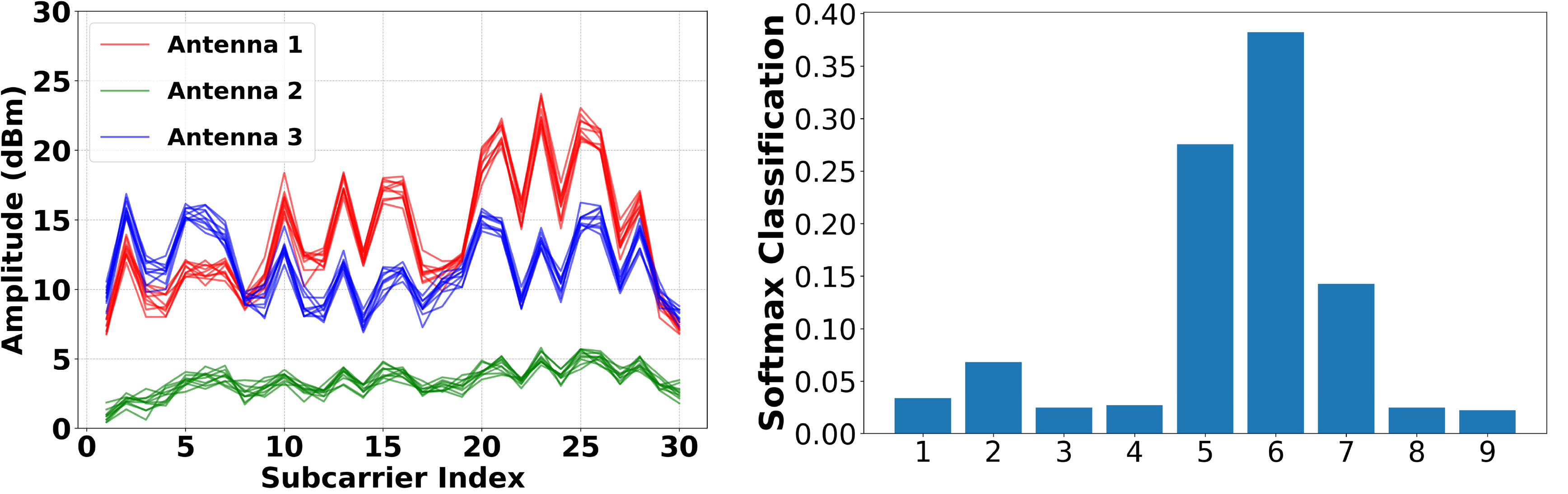}}
	\caption{Illustration of DNN-based location determination. (a) The deployment of the selected RPs, (b) Respective CSI amplitude distributions from the selected 9 RPs,   
		(c) Target location prediction with the softmax classification. }
	\label{fig_obs}
\end{figure}

\begin{figure*}[thbp]
	\centering
	\includegraphics[width=0.9\textwidth]{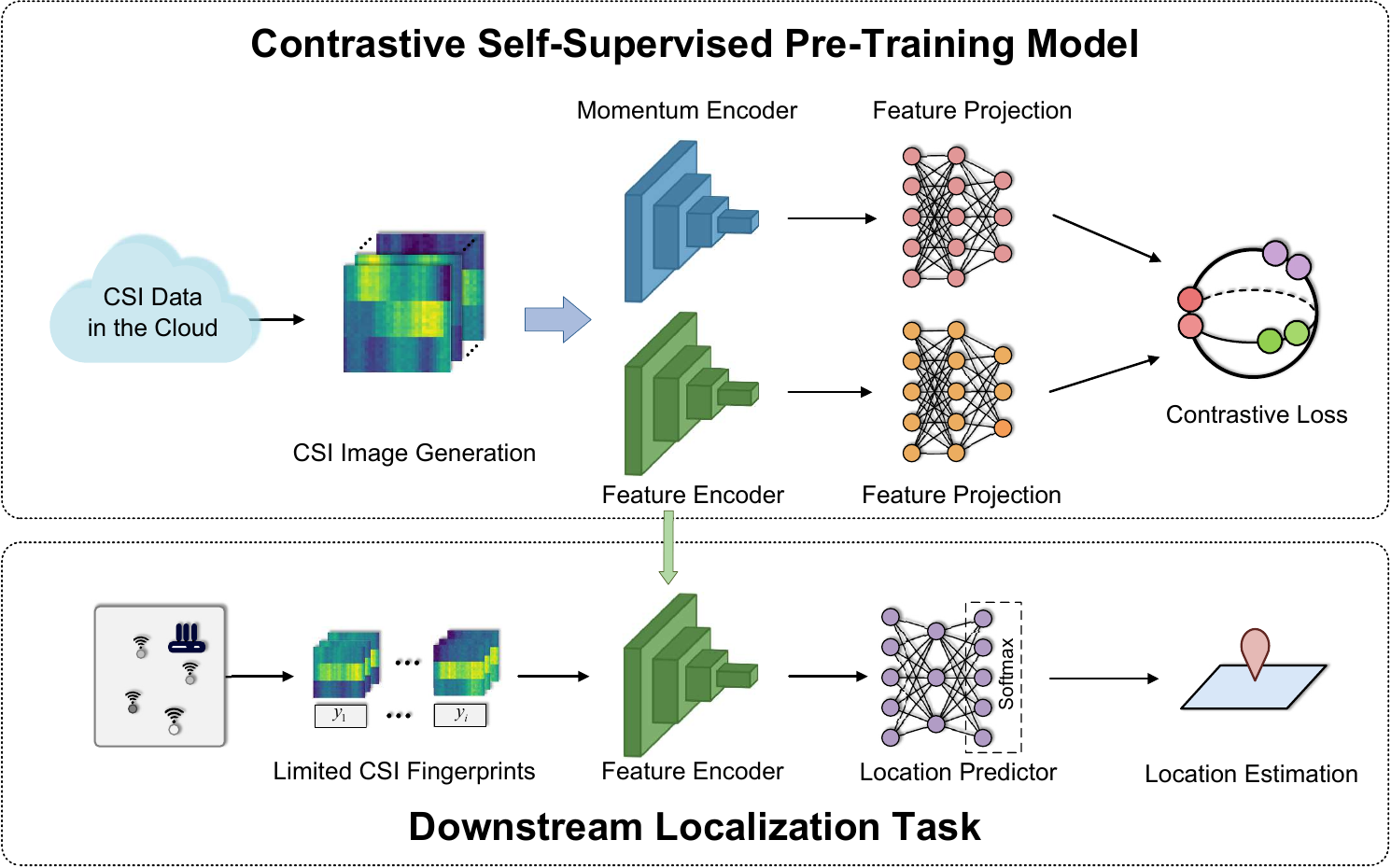}
	\caption{CSSLoc's Overview.}
	\label{fig_css}
\end{figure*}

In order to learn effective representations for cross-scenario localization at lower deployment and maintenance costs, we revisit the DNN-based localization process in the supervised learning method. 
Fig. 1 shows the experimental visualization with the location determination. In practice, several RPs are placed in a line with the different spacings by illustrated in Fig. 1(a), and the respective CSI amplitude distributions from the selected 9 RPs are plotted in Fig. 1(b) and fed into DNN to extract features. We use these CSI fingerprints to train the fingerprint-based localization model by the feature extractor with two full-connected layers and the location predictor with the softmax classification.  
In the localization stage, the CSI query from the neighboring RP 6 is received to estimate its location with the softmax output. As the illustration of the location determination in Fig. 1(c),
the location prediction is more likely to be RP 6, RP 5, or RP 7 with relatively high probabilities. 
Real-world observation shows that CSI distributions from the neighborhood RPs, such as RP 5, RP 6, and RP 7, have relatively similar fluctuations, whereas they are dissimilar from others. 
This resultant experiment uncovers that the learned feature representations can capture the characteristics of the apparent similarity and dissimilarity from radio data themselves, not from the location annotations \cite{InstDisc,InstSpread}. 
Therefore, it inspires us to learn an insightful metric on the similarity of radio samples in the representation space, without the location supervision, in an environment/scenario-agnostic way. 
We resort to contrastive self-supervised learning \cite{SimCRL,MoCo} to learn the discriminable representations so that the similar samples are closely clustered together and the dissimilar ones are separated. 
The learned representations can be directly transferred to downstream localization tasks in various application scenarios, pushing deep-learning-based localization from specificity to generality.

\section{System Design and Implementation}

CSSLoc proposes a generic framework of deep-learning-based localization with 
contrastive self-supervised learning to achieve accurate cross-scenario localization with limited deployment efforts. Without location information supervision, we focus on learning to discriminate the similarity and dissimilarity of annotated CSI samples in the representation space. Siamese neural networks  are developed to build up the contrastive self-supervised pre-training model for learning effective representations with the robustness of environment/scenario dynamics. 
For downstream
localization tasks, the learned feature encoder is directly
transferred to capture the appropriate features, and the location
predictor is trained with limited CSI fingerprints in an indoor
scenario, effectively reducing the deployment efforts.
As demonstrated in Fig. \ref{fig_css},  the proposed CSSLoc prototype adopts a cloud-edge implementation architecture \cite{Cosmo} which is elaborated in detail as follows.

The first stage is to build a \textbf{contrastive self-supervised pre-training model} with large-scale  CSI data centralized in the cloud.  We use CSI samples with multi-antenna and multi-subcarrier measurements to generate radio images and create the pre-training dataset with the defined positive and negative pairs from the corresponding same RP and the different RPs, respectively. 
Then the \textit{feature encoder} is designed with CNN architecture for extracting the feature vectors, and the \textit{feature projection} is further designed to perform a learnable nonlinear transformation between the feature vectors and the contrastive loss, which effectively promotes the quality of the learned feature representations for downstream tasks \cite{MoCo,MoCov2}. In the representation space, the \textit{contrastive loss} is formulated to 
enable the query to match with its positive key and separate from other channel keys. 
During the contrastive self-supervised pre-training, the \textit{momentum encoder} adopts a moving-averaged update to optimize the feature encoder, which can keep the similarity comparison as consistent as possible to learn effective representations for downstream localization tasks. 


The second stage on the edge is to perform the \textbf{downstream localization task}, The linear \textit{location predictor} is trained by limited CSI fingerprints with the location information. We freeze the trained feature encoder to directly extract the appropriate features for robust and accurate location estimations at lower deployment and maintenance efforts. When real-time CSI measurements are received online to generate radio images in a specific scenario, the current features are captured to estimate the target locations. 
Finally, the \textit{location estimations} are determined with the softmax outputs in a probabilistic method. 

\subsection{CSI Image Generation}
 
In contrastive self-supervised learning framework, classical methods \cite{MoCo,SimCRL,InstDisc} rely on diverse data augmentation of an image instance, such as random cropping, color distortion, Gaussian blur, etc. However, we cannot directly apply these image-data-specific augmentations to perform radio data processing, because these image augmentations would alter the information carried by radio signals. In this case, we fully utilize the inherent consistencies of CSI samples from the same RP, as the positive pairs, and their discrepancies of different RPs, as the negative pairs, to create the pre-training dataset  for CSSLoc's contrastive self-supervised learning.   

\begin{figure}[tbp]
	\centering
	\includegraphics[width=0.36\textwidth]{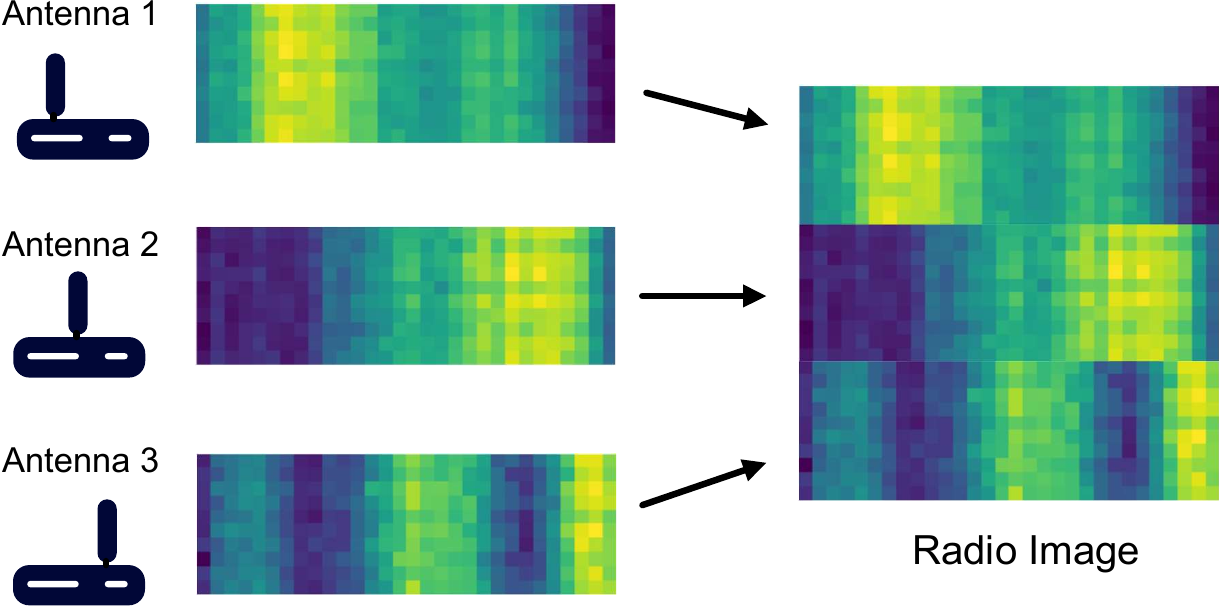}
	\caption{CSI  image generation.}
	\label{radio image}
\end{figure}

Furthermore, for learning finer representations for downstream tasks in various scenarios, we convert CSI samples into radio images to augment the order of magnitude of pre-training data  \cite{DadLoc,ILOT}. 
With multi-antenna and multi-subcarrier radio measurements \cite{tool2011},
tens of CSI samples are collected to generate CSI radio images. 
These CSI amplitudes at each antenna are transformed into an individual radio image, and these radio images from an antenna array are stitched into one CSI image. We demonstrate the data processing of CSI image generation in Fig. \ref{radio image}. 

The generated CSI images are from the same RP with the inherent similarity, definitely as the positive pairs. Meanwhile, the dissimilar CSI images at different RPs are the negative pairs. 
We feed these CSI images into the developed Siamese neural networks, and CSSLoc attempts to learn the meaningful representations that uncover their similarity and dissimilarity without location information supervision in complicated indoor environments.


\subsection{Contrastive Self-supervised Learning}




In contrastive self-supervised learning framework, CSSLoc develops Siamese network structure to learn essential representations from CSI data with their similarity discrimination.   
We employ CNN architecture to capture the feature vectors from the generated CSI images and exploit the low-dimensional projection to improve the quality of the feature representations for downstream localization tasks \cite{SimCRL,MoCo}. 
The contrastive loss is formulated to make the similar embeddings closer and push the dissimilar embeddings away in the representation space \cite{MoCo,InstDisc}.
The detailed implementation of our contrastive self-supervised pre-training is elaborated as follows. 

\subsubsection{\textbf{Feature Encoder}}
By feeding unannotated CSI images, 
the feature encoder $f$ is designed with the CNN architecture 
to extract the feature vectors as 
\begin{equation}
    \mathbf{v} = f(\mathbf{x})=\operatorname{CNN}(\mathbf{x}),
\end{equation}
where $\mathbf{x}$ is the input of generated CSI images and the feature vectors $\mathbf{v}$ are the output of the feature encoder.
Our CNN architecture includes one convolutional layer and one max-pooling layer. 
CSSLoc leverages such two CNN structures to design the feature encoder which can be directly used to capture the feature vectors for downstream tasks, once trained with network convergence.


\subsubsection{\textbf{Feature Projection}}

The projection networks $g(\mathbf{v})$ are trained to map the feature vectors into the latent low-dimensional space \cite{SimCRL}. We adopt two multilayer perceptron (MLP) to improve the quality of the learned feature representations \cite{MoCo,SimCRL}. Two MLPs with ReLU activations are designed to perform a learnable nonlinear transformation as
\begin{equation}
    \mathbf{z} = g(\mathbf{v})=\operatorname{MLP}(f(\mathbf{x})). 
\end{equation}

For the similarity discrimination, CSSLoc designs the twin-branch projection networks to obtain the low-dimensional embeddings  with contrastive self-supervised learning. 
The query $\mathbf{z}_0=g(f_e(\mathbf{x}_0))$ is denoted as the query CSI image $\mathbf{x}_0$ in one-branch projection network. Meanwhile, the embeddings, as the channel keys, are learned as $\mathbf{z}_k=g(f_k(\mathbf{x}_k)), k=1, 2, \ldots, K$ in the other-branch network. They are illustrated by Fig. \ref{NetAchi} in the next section.

\subsubsection{\textbf{Contrastive Loss}}
The contrastive loss is formulated to carry out the similarity discrimination between the feature embeddings in the latent embedding space.  
We store an embedding query $\mathbf{z}_0$ and the encoded embeddings $\{\mathbf{z}_1, \mathbf{z}_2 \ldots,\mathbf{z}_K\}$ as the given channel keys.  The query $\mathbf{z}_0$ should be matched with its positive key $\mathbf{z}_{+}$, since they are the positive pair with the similar CSI images from the same RP, at the same time, the negative pairs are consisted of $\mathbf{z}_0$ and the others.
The contrastive loss \cite{MoCo,SimCRL,NCE} is formulated to 
learn the inherent similarities and dissimilarities between positive and negative pairs in the representation space.

Specifically, the Cosine similarity is adopted to quantify the similarity between different encoded embeddings with $\ell_2$ normalization as
\begin{equation}
\mathrm{sim}(\mathbf{z}_i,\mathbf{z}_j)=\frac{\mathbf{z}_i^{\mathrm{T}}\mathbf{z}_j}{\left\|\mathbf{z}_i\right\|\left\|\mathbf{z}_j\right\|}.
\end{equation}
The contrastive loss \cite{MoCo,DeepInfo} is a function of identifying a positive pair with a lower value and separating other negative pairs away, which is denoted by one positive pair and $K$ negative samples as 
\begin{equation}
        \mathcal{L}_{o}= -\log \frac{\exp{\left(\mathrm{sim}(\mathbf{z}_0,\mathbf{z}_{+}) / \tau\right)}}{\sum_{k=0}^{K} \exp{\left(\mathrm{sim}(\mathbf{z}_0,\mathbf{z}_{k}) / \tau\right)}},
\end{equation}
where $\tau$ is a temperature parameter \cite{SimCRL,MoCo} that controls the sensitivity of the similarity metric between different embeddings. $\mathcal{L}_o$, defined as InfoNCE loss \cite{NCE,DeepInfo,MoCo}, is the log loss of a ($K$+1)-way softmax classification that indicates to classify $\mathbf{z}_0$ as $\mathbf{z}_{+}$.  

In our contrastive self-supervised pre-training, we set multiple positive pairs to enhance the effectiveness of the learned feature representations.
The query $\mathbf{z}_i$ has the positive pair $\mathbf{z}_+$ and its negative pairs $\mathbf{z}_k, k=1,2,\ldots,K$. the  the contrastive loss  can be expressed as
\begin{equation}
        \mathcal{L}= \mathbb{E}_{\{\mathbf{x_i},\mathbf{x_+}\},\{\mathbf{x}_{k}\}}\left[-\log \frac{\exp{\left(\mathrm{sim}(\mathbf{z}_i,\mathbf{z}_{+}) / \tau\right)}}{\sum_{k=0}^{K} \exp{\left(\mathrm{sim}(\mathbf{z}_i,\mathbf{z}_{k}) / \tau\right)}}\right].
\end{equation}
For learning fine representations, the channel keys should be large enough to cover the abundant negative pairs. 
In parallel,  CSSLoc should keep the channel keys as consistent as possible to perform the similarity metrics during network optimization by back-propagation \cite{MoCo,MoCov2}. 

\subsubsection{\textbf{Training Strategy}}
CSSLoc learns meaningful representations through the contrastive loss which enable the similar embeddings to cluster together and separate the dissimilar embeddings away.  
It involves random sampling to calculate the contrastive loss and updating the feature encoder to maintain the key representation's consistency with the similarity comparison when optimizing the developed Siamese neural networks \cite{MoCo}.

\textit{Random Sampling}: We select all CSI images from the training set to feed into the developed Siamese neural networks for contrastive self-supervised learning. 
All channel keys stored in a queue contain the positive pairs and a large number of negative pairs. In one batch, the CSI samples are randomly selected to form multiple positive and negative pairs. The contrastive loss is computed by Eq. (6) to optimize the learnable parameters of the developed DNNs by back-propagation. 

\textit{Momentum Update}: 
\label{MomentumMechanism}
It is intractable to optimize the feature encoder by back-propagation which should propagate the gradients to all CSI images. Only the feature encoder on the query sample is updated, and then its hyper-parameters are copied to the key encoder, which the mutated feature encoder reduces the consistency of embedding keys' representations to yield poor performance \cite{MoCo}. 

We adopt a moving-averaged update, as the \textit{momentum encoder}, to optimize the DNNs for learning good representations. The momentum encoder $f_k$ is updated with the learnable parameters $\xi$ as
\begin{equation}
    \xi \leftarrow m \xi+(1-m) \theta,
\end{equation}
where $m\in [0,1)$ is defined as a momentum coefficient. $\theta$ is the learnable parameters of the feature encoder $f_e$ and only $\theta$ are optimized by back-propagation. 

\begin{figure}[tbp]
     \centering
     \includegraphics[width=0.8\linewidth]{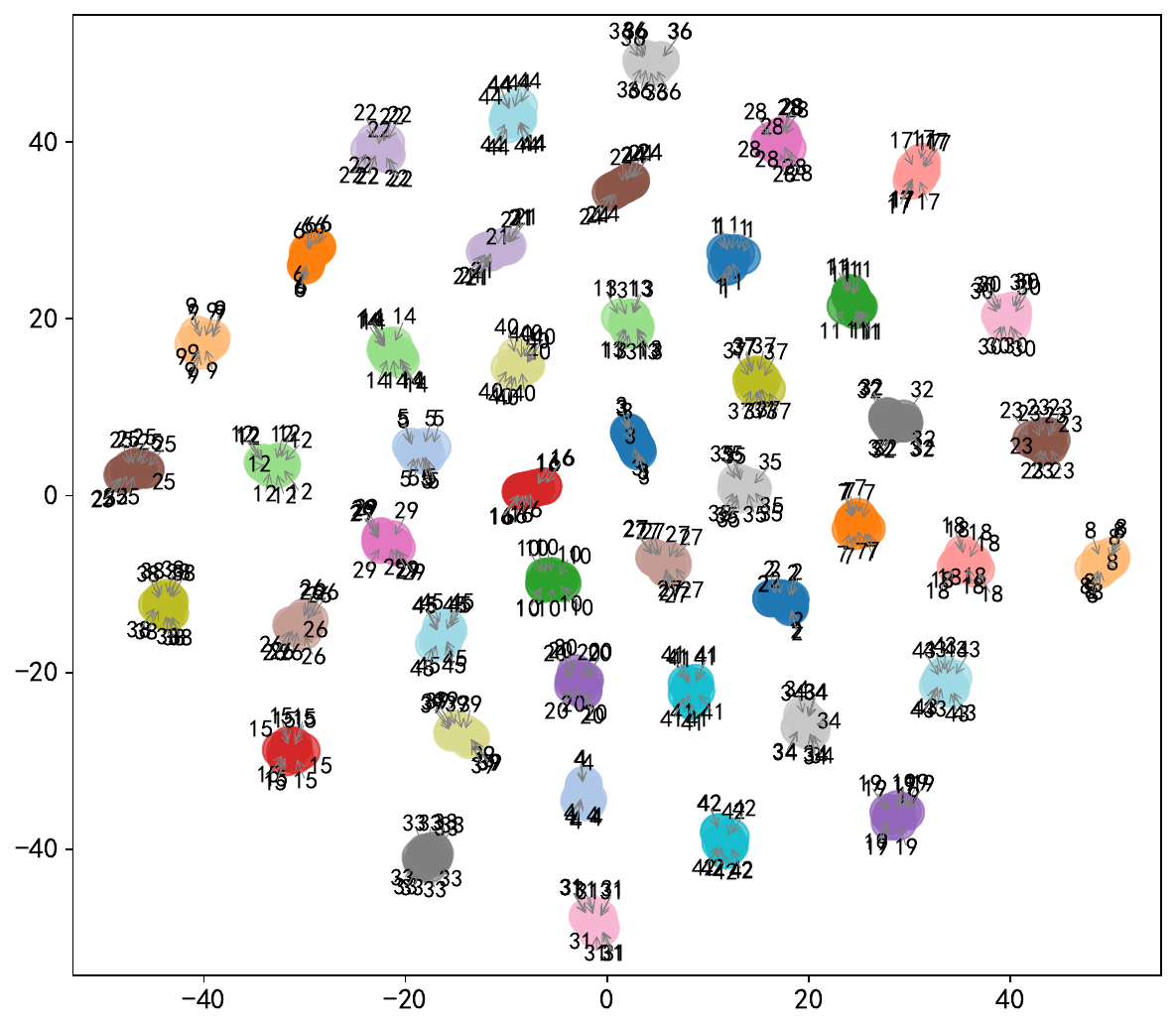}
     \caption{Illustration of CSSLoc's learned feature representations.}
     \label{tsne}
 \end{figure}


We demonstrate the informative visualization of learned feature representations with t-SNE analysis in Fig. \ref{tsne}. Without location information supervision, CSSLoc can learn discriminative representations which can not only cluster the features of similar CSI images with the same color,  but also separate the features from the dissimilar ones with different colors to indicate different feature representations. 
Furthermore, the details of CSSLoc's pre-training model with contrastive self-supervised learning is presented in Algorithm 1.

\begin{algorithm}[t]
	\renewcommand{\algorithmicrequire}{\textbf{Input:}}
	\renewcommand{\algorithmicensure}{\textbf{Output:}}
	\removelatexerror
\caption{Contrastive Self-supervised Learning for CSSLoc's pre-training model}
\label{alg:csl}
\begin{algorithmic}[1]
\REQUIRE Unannotated CSI Images $\mathcal{X} = \{\mathbf{x}_i\}_{i=1}^M$, the batch size $B$, the queue size $N$, the temperature coefficient $\tau$, and  the momentum coefficient $m$. 
\ENSURE The feature encoder $f_e$ for downstream localization task.

\STATE \textbf{Initialize}: 
The parameters of the feature encoder $\theta_e$ and the momentum encoder $\theta_k \gets \theta_e$, the parameters of the respective projection networks $\phi_e, \phi_k$, and the FIFO queue $\mathcal{Q}$ with the $K$ channel keys. 
\FOR{each training iteration}
    \STATE Identify the positive pairs of $\mathbf{x}_i$ and $\mathbf{x}_+$ and the corresponding negative pairs  $\mathbf{x}_k$
    \STATE Capture the feature vectors $\mathbf{v}_i=f_e(\mathbf{x}_i)$ and $\mathbf{v}_k=f_k(\mathbf{x})$ in two-branch networks
    \STATE Obtain the queries $\mathbf{z}_i = g(f_e(\mathbf{x}_i))$ and the channel keys $\mathbf{z}_k = g(f_k(\mathbf{x}))$
    \STATE Enqueue current key $\mathcal{Q} \gets \mathcal{Q} \cup \{\mathbf{z_k}\}_{i=1}^B$
    \STATE Compute the contrastive loss with Eq. (6)
    
    \STATE Update $\theta$ via $\nabla_{\theta} \mathcal{L}$ 
    \STATE Momentum update by Eq. (7)
\ENDFOR
\STATE \textbf{Return} $f_e$
\end{algorithmic}
\end{algorithm}

\subsection{Downstream Localization Task}
For downstream localization tasks, the fixed feature encoder $f_e$ is directly transferred to capture the feature representations underlying CSI fingerprints in various  indoor scenarios. 
CSSLoc uses the limited  CSI fingerprints $\mathcal{D}=\{(\mathbf{x}_i,y_i)\}_{i=1}^{N}$ from $r$ RPs to train the location  predictor $p$ with softmax classification \cite{DadLoc,l2l}.  The training loss is written as
\begin{equation}
    \mathcal{L}_p=\mathbb{E}_{(\mathbf{x}_i,y_i)\in \mathcal{D}} \ell_{ce} \left(p(f_e(\mathbf{x}_i)),y_i\right),
\end{equation}
where $\ell_{ce}$ is a cross-entropy loss.  the output of the location predictor is $\hat{y}=p(f_e(\mathbf{x}_i))$.

When receiving CSI measurements convert into CSI image $\hat{\mathbf{x}}$ in a specific scenario, the feature vectors $\mathbf{v}=f_e(\hat{\mathbf{x}})$ can be obtained by the frozen feature encoder $f_e$ from the contrastive self-supervised pre-training model. 
With the softmax output, 
the posterior probability of the location predictor $Pr(y|\mathbf{v})$ is calculated to determine the location estimation $\hat{L}$ as
\begin{equation}
    \hat{L}=\sum_{i=1}^{r} Pr(y|\mathbf{v}) L_y,
\end{equation}
where $L_y$ is the location information with the corresponding location label $y$ at the $r$th RP, and $r \ll R$.

\begin{figure*}[thbp]
	\centering
        \subfigure[Corridor]{
		\includegraphics[width=0.35\textwidth]{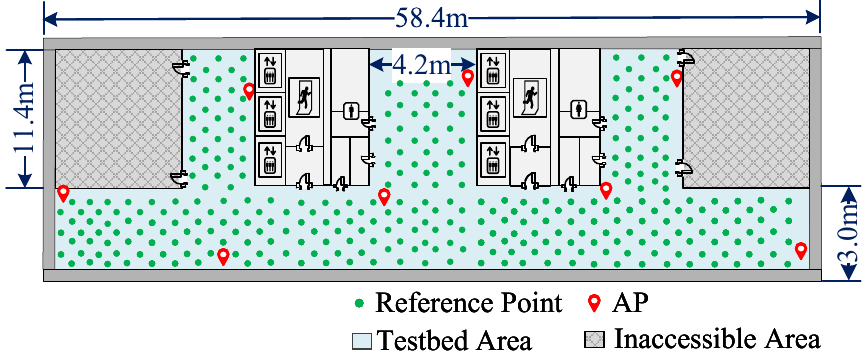}} \quad 
	\subfigure[Hall]{
		\includegraphics[width=0.22\textwidth]{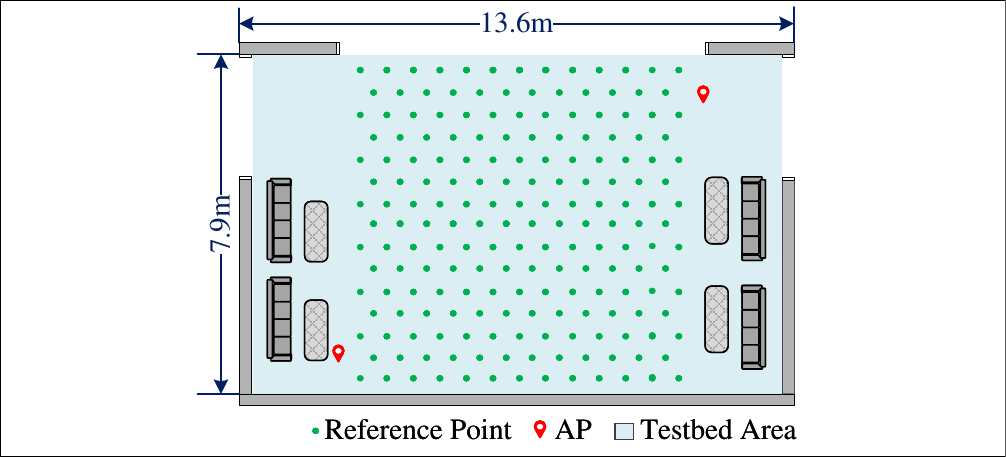}} \quad 
	\subfigure[Lounge]{
		\includegraphics[width=0.32\textwidth]{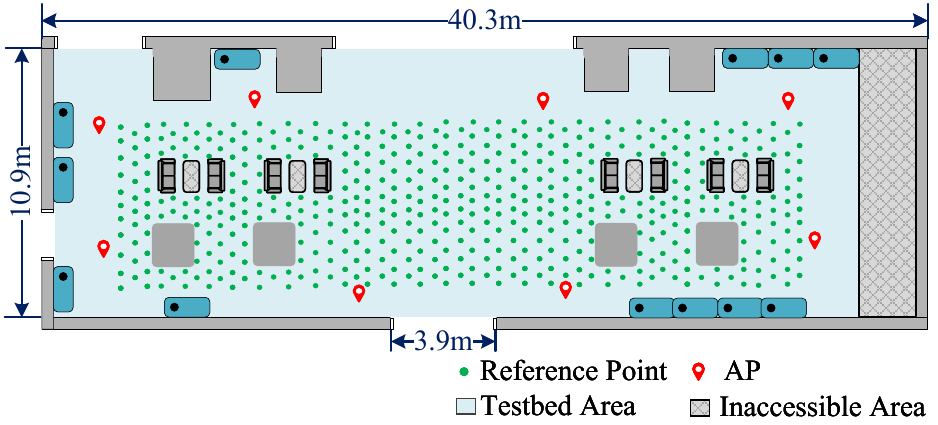}}
	\caption{Floor layouts of indoor scenarios. }
	\label{fig_floorplan}
\end{figure*}

\section{Experimental Evaluation}

In this section, we introduce the experimental settings and methodology, then the performance evaluation and analysis on the proposed CSSLoc system are presented with extensive real-world experiments in diverse indoor scenarios. 

\subsection{Experiment Methodology}


\subsubsection{CSI Collection}
We extensively collect CSI data by using off-the-shelf WiFi infrastructures with commercial 802.11n 5300 NICs \cite{tool2011}. The WiFi transceivers are set in the injection mode with MIMO-OFDM modulation scheme in the 5.32GHz spectrum (CH64) by the sampling frequency of 100 Hz. 
The practical experiments are conducted on the typical indoor scenarios as shown in Fig. \ref{fig_floorplan}. 
In every scenario, we deploy the pre-defined RPs to measure CSI readings for training our self-supervised pre-training model and record their corresponding location information of all test location points to evaluate their localization accuracy. In this case, the generated CSI images from the same RP own the intrinsic consistency to become multiple positive pairs, whereas CSI images from different RPs become the negative pairs.  CSSLoc adopts contrastive self-supervised learning to capture the meaningful representations which can characterize the similarities and dissimilarities of the unannotated CSI images. The learned representations are directly transferred to realize downstream robust localization tasks in various scenarios,  pushing DNN-based localization from specificity to generality for offering fully practical LBS.

\subsubsection{Datasets}
The training dataset consists of CSI data from three typical indoor scenarios, and their CSI images are generated to learn the contrastive self-supervised pre-training model in a scenario-agnostic manner. 
We also prepare the testing dataset from the individual indoor scenario to evaluate localization performance with extensive experiment study, in order to validate the effectiveness of the proposed CSSLoc system.

The first testbed area,  the \textit{corridor}  in Fig. \ref{fig_floorplan}(a), is the size of 298m$^2$ with dense multipath propagation. 
We randomly select 359 RPs with CSI data for the training dataset. The number of testing points is 360 in the accessible area.
The second scenario is the \textit{hall}, the size of 107m$^2$, under much light-of-sight (LOS) propagation. There are 188 RPs deployed as the training location points and 187 testing location points for the downstream localization tasks. 
The last scenario as the \textit{lounge} with complicated wireless propagation is the size of  
436 m$^2$.  The furniture and obstacles cause non-LOS (NLOS) propagation. We set 425 RPs with corresponding CSI measurements for training our wireless pre-training model and 426 RPs for evaluating the localization accuracy of the proposed CSSLoc prototype.

\subsubsection{Network Architectures}
\begin{figure*}[th]
	\centering
		\includegraphics[width=0.9\linewidth]{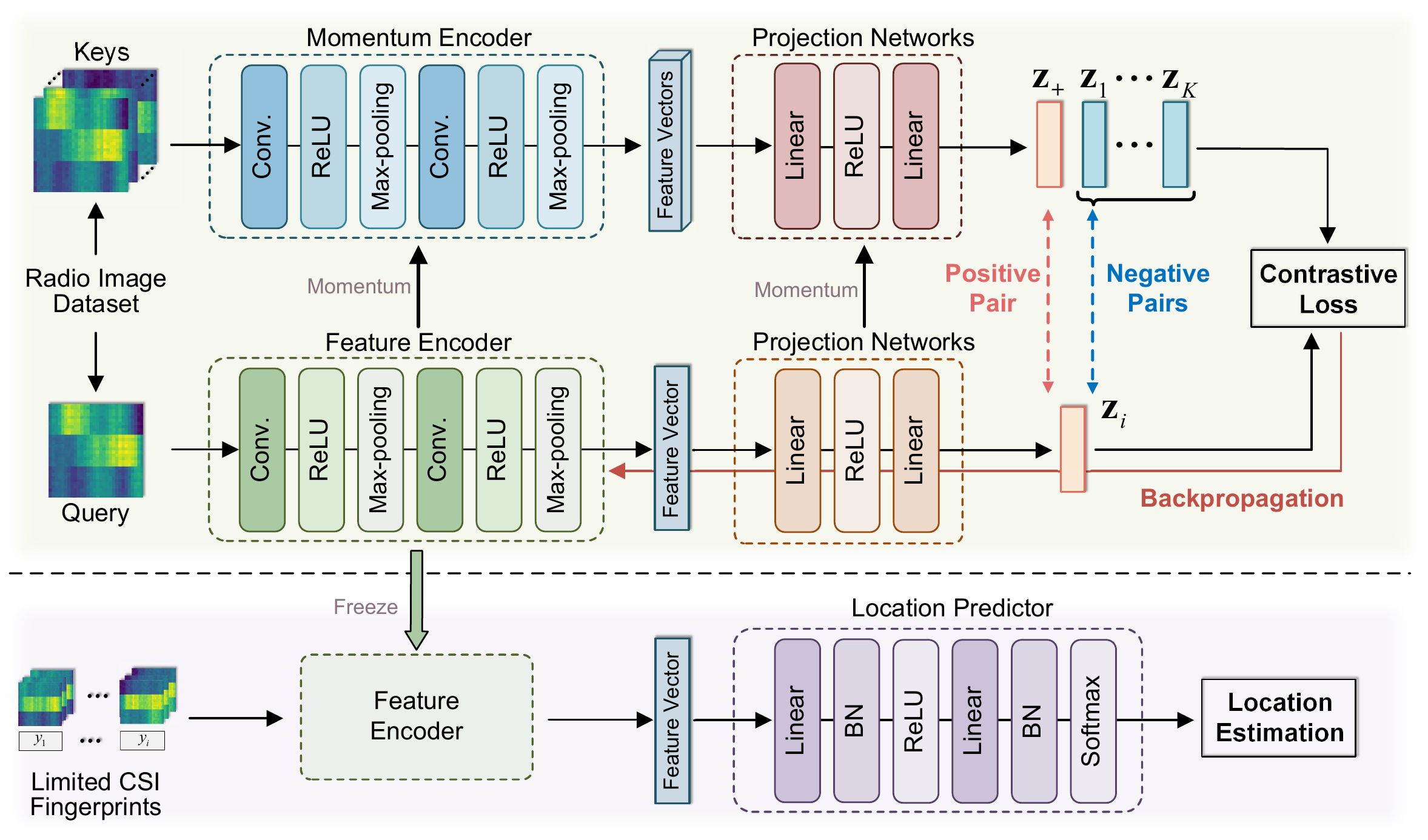}
	\caption{CSSLoc's Network Architectures.}
	\label{NetAchi}
\end{figure*}

As demonstrated in Fig. \ref{NetAchi}, CSSLoc's architecture involves the \textit{contrastive self-supervised pre-training model} with the feature encoder, the momentum encoder, the projection networks, and \textit{downstream localization task} with the frozen feature encoder and the location predictor. The details are introduced as follows.

In the contrastive self-supervised pre-training model, the feature encoder utilizes two convolutional structures including one convolutional layer and one max-pooling layer. 
 The size of one CSI radio image is 30x30. The first convolutional layer uses 4 kernels with the size of 3x3, the channel numbers of $\{1, 4, 4, 4\}$, a stride of 1, and a padding of 1. The second convolutional layer has 4 kernels with a size of 3x3, a stride of 1, and a padding of 1. The ReLU non-linearity and the max-pooling are applied to extract the feature vectors with the dimension of 100. The momentum encoder owns the same architecture as the feature encoder with the momentum update. 
 For the projection networks, there are two fully connected (FC) layers by ReLU activation with 100 and 32 neurons in each layer, respectively. 
For downstream localization tasks, the location predictor is designed by two FC layers $\{100, 32\}$ with ReLU activation and the softmax layer with the number of RPs in a specific indoor scenario for location prediction. 

\subsubsection{Default Settings}

One CSI image is generated with the amplitude distributions of 10 CSI samples from 30 subcarriers at 3 antennas. All CSI images from the pre-training set are fed into the developed Siamese nerual networks, and 
we train the contrastive self-supervised pre-training model with a batch size 1024 in the 150 epochs. The contrastive loss is formulated with the similarity metric with the temperature coefficient of 0.03. 
We optimize the feature encoder and the corresponding projection networks by using Adam optimizer with the learning rate of $5 \times 10^{-3}$ and the weight decay of $5 \times 10^{-4}$. The momentum encoder is updated by a moving-averaged moment coefficient with 0.99.  For downstream localization tasks, CSSLoc can only conduct half of the site survey to train the location predictor and achieve satisfactory localization performance, effectively reducing system deployment efforts. 

\subsubsection{Comparative Approaches}
In order to validate its extensive effectiveness, we compare the localization performance of the CSSLoc's prototype with classical and state-of-the-art deep-learning-based schemes:
\begin{itemize}
 \item  Supervised learning-based methods: WiDeep \cite{WiDeep},   CNNLoc \cite{CNNLoc}, and GNN-Loc \cite{GNN-Loc}; 
\item 
Semi-supervised learning-based method: ReNet-Loc with relation networks \cite{l2l}; 
\item Unsupervised learning-based method: DeepFi \cite{DeepFi}.  \end{itemize}

\subsection{Performance Evaluation and Discussion}

\subsubsection{Accuracy Performance}
\begin{figure}[th]
	\centering
		\includegraphics[width=0.9\linewidth]{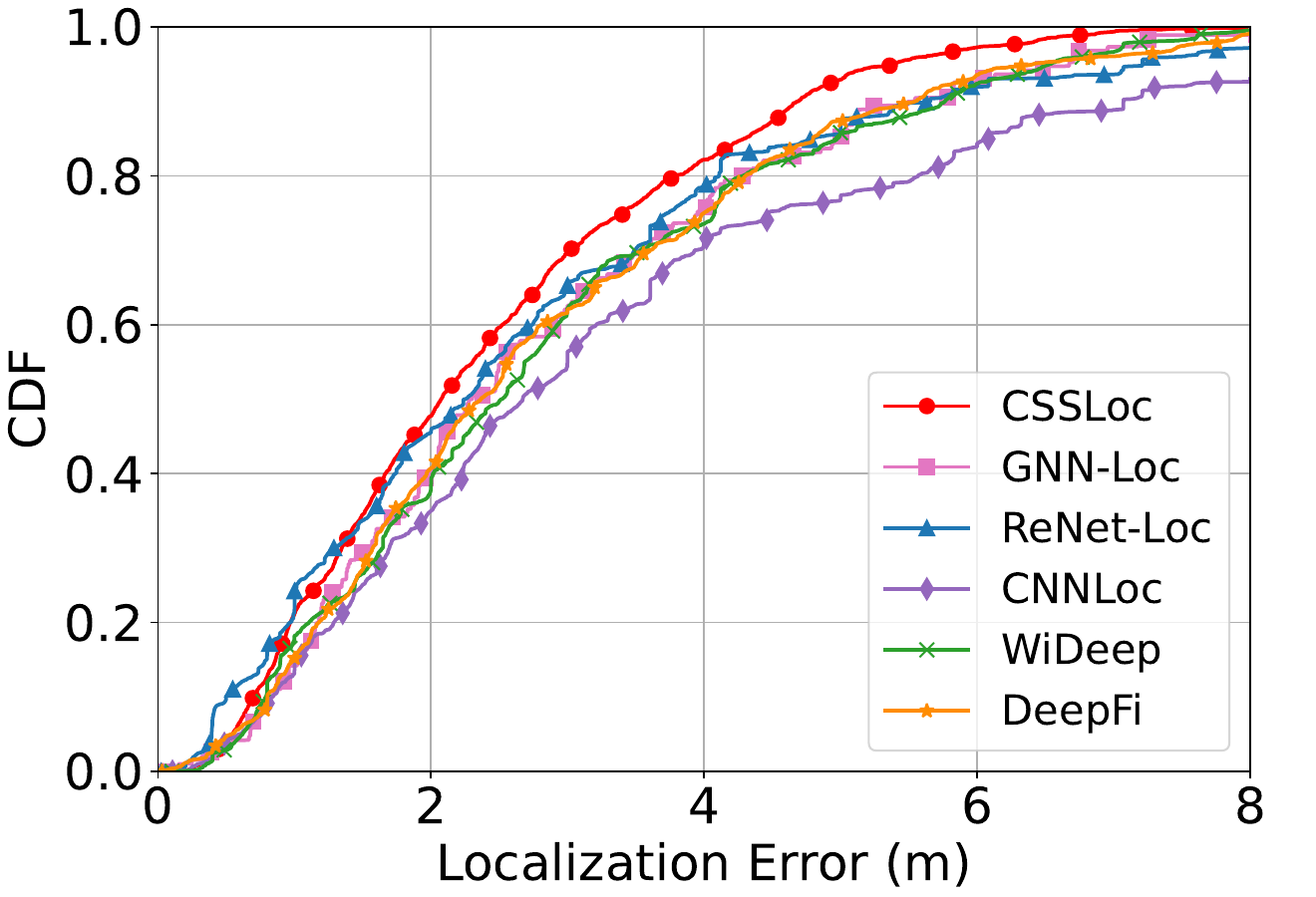}
	\caption{Overall localization accuracy comparison.}
	\label{over-cdf}
\end{figure}

We first calculate and compare the localization accuracy of the proposed CSSLoc system against others in these typical indoor scenarios.  As illustrated in Fig. \ref{over-cdf}, the cumulative distribution functions (CDF) of localization errors are shown by the root mean square errors (RMSE) of the proposed CSSLoc and other comparative schemes. CSSLoc can achieve the median error of 1.97m with an accuracy improvement of 27.1\% for CNNLoc,  21.5\% for WiDeep, 13.4\% for GNN-Loc, respectively. 
For the 80th percentile accuracy,  CSSLoc can outperform 30.2\% for CNNLoc, 7.7\% for WiDeep, and 7.7\% for GNN-Loc. These localization results can demonstrate that CSSLoc can outperform its supervised representation counterpart with contrastive self-supervised representation learning. Furthermore, CSSLoc can show the median accuracy improvement of  16.6\%  for DeepFi with unsupervised representation learning. 
We also compare the localization accuracy with metric learning in a semi-supervised manner as ReNet-Loc that CSSLoc can gain 12.1\% median accuracy improvement and 0.16m for the 80th percentile accuracy, respectively. 



TABLE I summarizes the localization performance compared with other evaluation indicators, such as the mean absolute error (MAE) and the standard deviation (STD).  With overall localization accuracy evaluation, the proposed CSSLoc scheme can reduce the localization error of  29.2\% for CNNLoc, 12.6\% for  WiDeep, 9.8\% for  GNN-Loc, 14.6\% for DeepFi, and 7.6\% for ReNet-Loc, respectively. Meanwhile, CSSLoc has achieved a lower STD of 1.68m than others, which can indicate that it can realize more robust location estimations than others in diverse indoor scenarios. 
With these extensive localization results, the developed CSSLoc prototype can achieve more satisfactory accuracy than others in complicated indoor environments, which benefits from the learned effective representation with contrastive self-supervised learning to acquire the environmental adaptability in complicated indoor scenarios.

\begin{table}[thbp]
	\renewcommand\arraystretch{1.25}
	\caption{Comparison of Localization Performance with different DNN-based models}
	\centering
	\label{comparetable}
	\begin{tabular}{|c|c|c|c|}
		\hline
		{Method}& RMSE      $\hat{\epsilon}\; (m)$ & MAE  $\bar{\epsilon} \; (m)$ & STD $\sigma \; (m)$   \\ \hline
		{CNNLoc}   & 4.31& 3.48&2.53\\ \hline
		{WiDeep}   & 3.49& 2.95&1.86\\ \hline
		{DeepFi}   & 3.57& 3.00&1.94\\ \hline
		{ReNet-Loc}   & 3.30& 2.66&1.96\\ \hline
        {GNN-Loc}   & 3.38& 2.87&1.79\\ \hline
		{\textbf{CSSLoc}}   &\textbf{3.05} &\textbf{2.55} &\textbf{1.68}\\ \hline               
	\end{tabular}
\end{table}

\subsubsection{Robustness Performance}
\begin{figure}[th]
	\centering
		\includegraphics[width=0.9\linewidth]{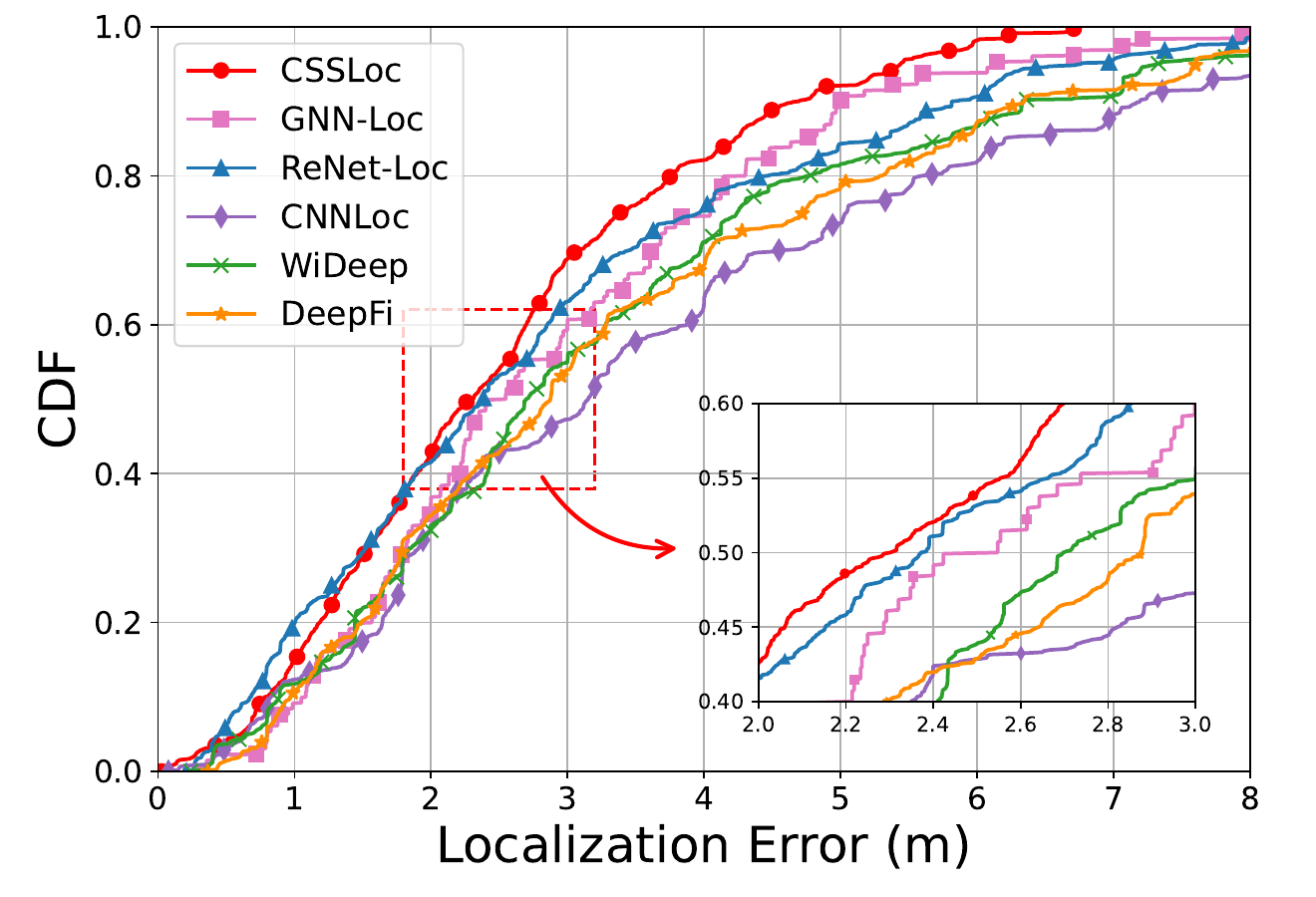}
	\caption{Robust localization performance comparison.}
	\label{robust-cdf}
\end{figure}

We evaluate the localization performance with the robustness of environmental dynamics. CSI fingerprints are collected at different periods with diverse environmental dynamics. 
For the performance evaluation of robust localization, both the learned feature encoder and the trained location predictor are frozen to estimate the locations in the downstream tasks. 
As localization results in Fig. \ref{robust-cdf}
CSSLoc also shows the superiority of localization accuracy over others with the median error of 2.29m, which improves the 27.4\% accuracy of CNNLoc, 15.2\% for WiDeep, 5.7\% for GNN-Loc, 3.6\% for ReNet-Loc, and 20.5\% for DeepFi, respectively. For the 80th percentile accuracy, CSSLoc also attains the accuracy improvement of 32.7\% for CNNLoc, 20.8\% for WiDeep, 9.4\% for GNN-Loc, 28.9\% for ReNet-Loc, and 12.9\% for DeepFi, respectively. 

Above all, these localization results can verify the effectiveness of the developed CSSLoc prototype for cross-scenario localization tasks in various indoor environments.   
The proposed framework of our CSSLoc system benefits from contrastive self-supervised learning to 
perform the CSI discriminate with the similarities and dissimilarities underlying their generated images in the representation space, without location information supervision.  In a scenario-agnostic manner, the proposed CSSLoc scheme is validated to learn generic representations which are directly transferred to realize accurate cross-scenario localization with satisfactory performances. 




\begin{figure*}[thbp]
\begin{minipage}{0.32\linewidth}
		\centering
		\includegraphics[width=0.98\textwidth]{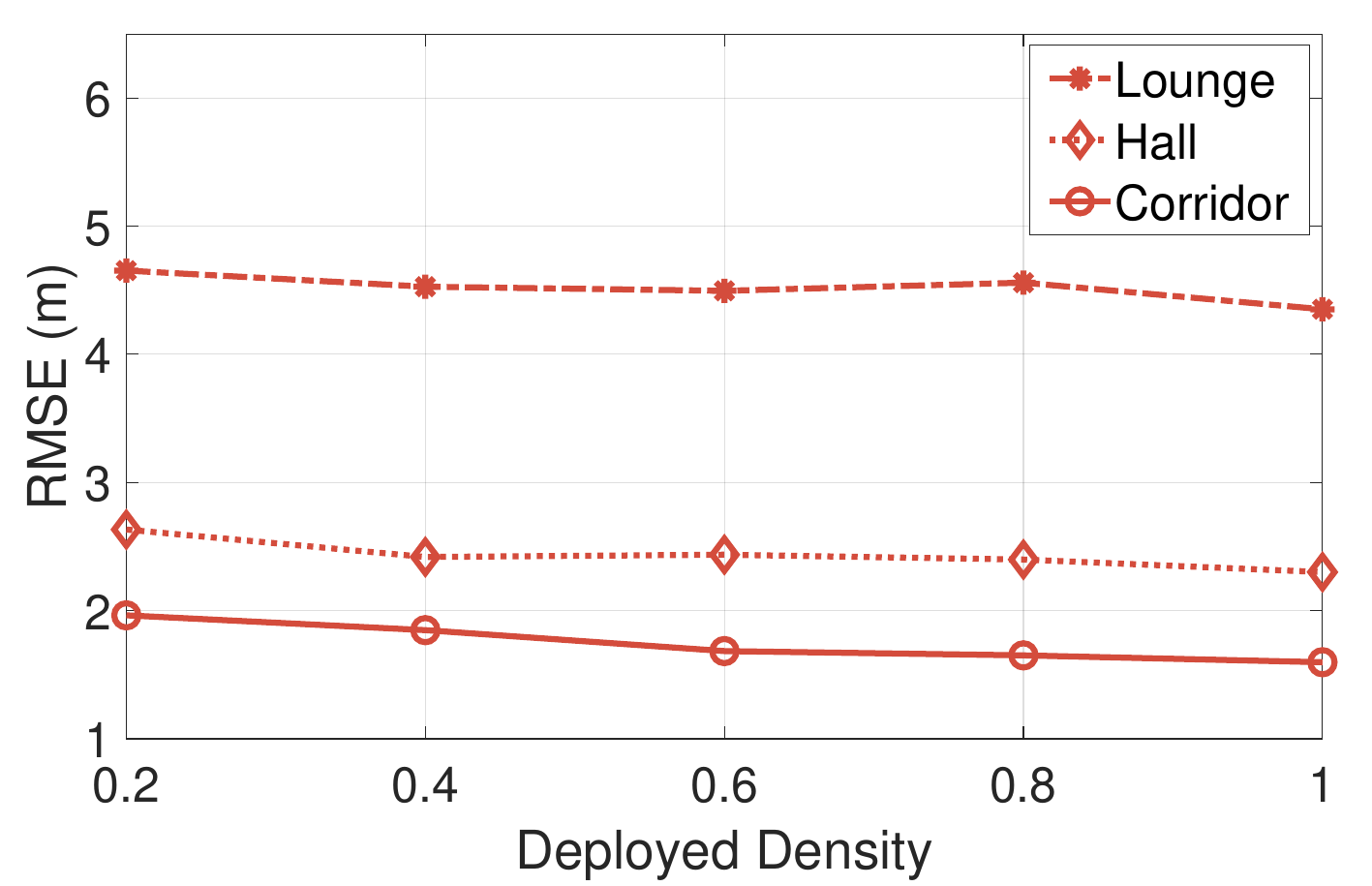}
		\caption{Localization Accuracy with the deployed density.}
		\label{deploy-dense}
	\end{minipage}
	\begin{minipage}{0.32\linewidth}
		\centering
		\includegraphics[width=0.96\textwidth]{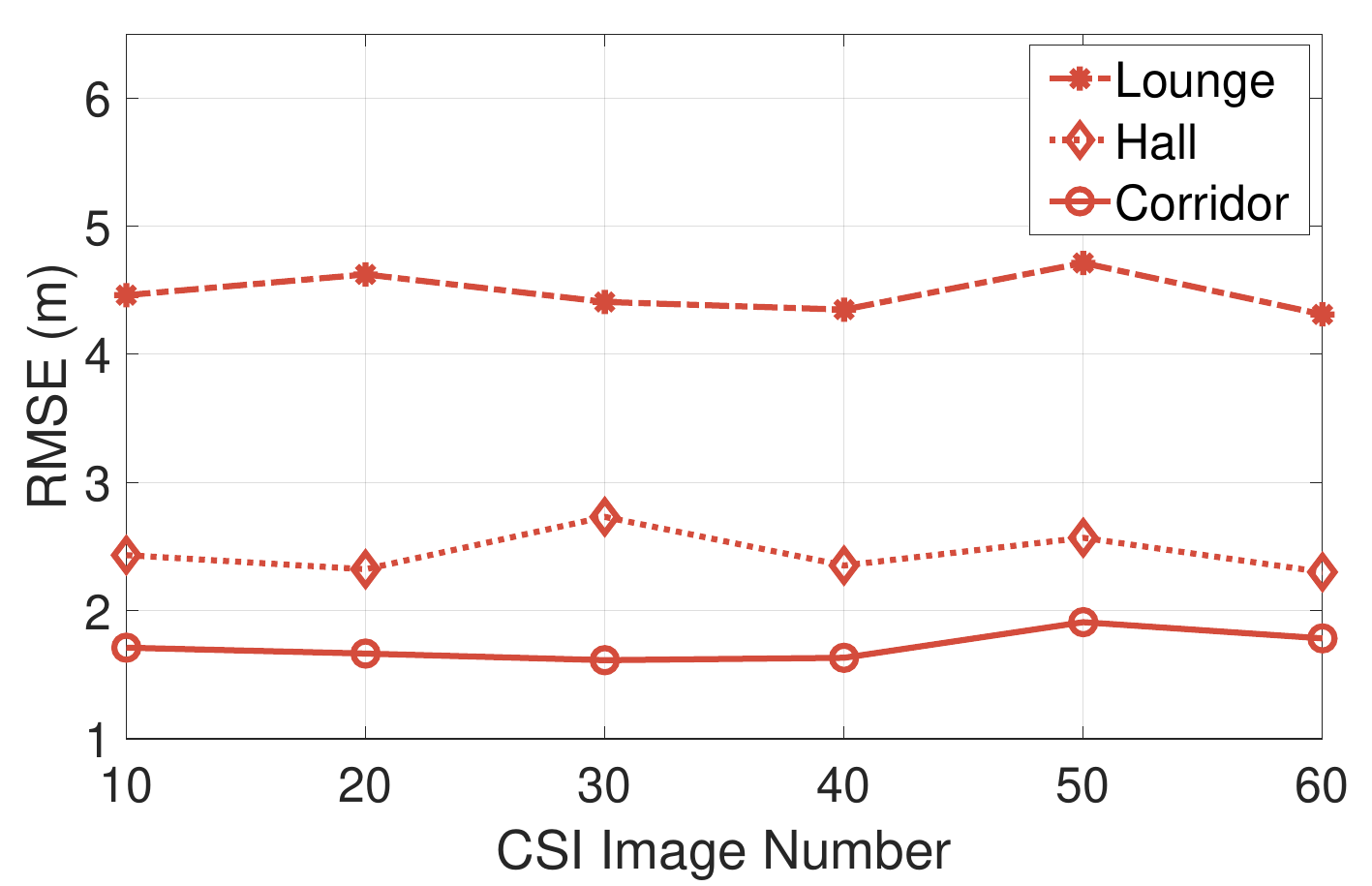}
		\caption{Localization accuracy with different CSI Image Numbers}
		\label{ImageSample}
	\end{minipage}
	\begin{minipage}{0.32\linewidth}
		\includegraphics[width=0.98\textwidth]{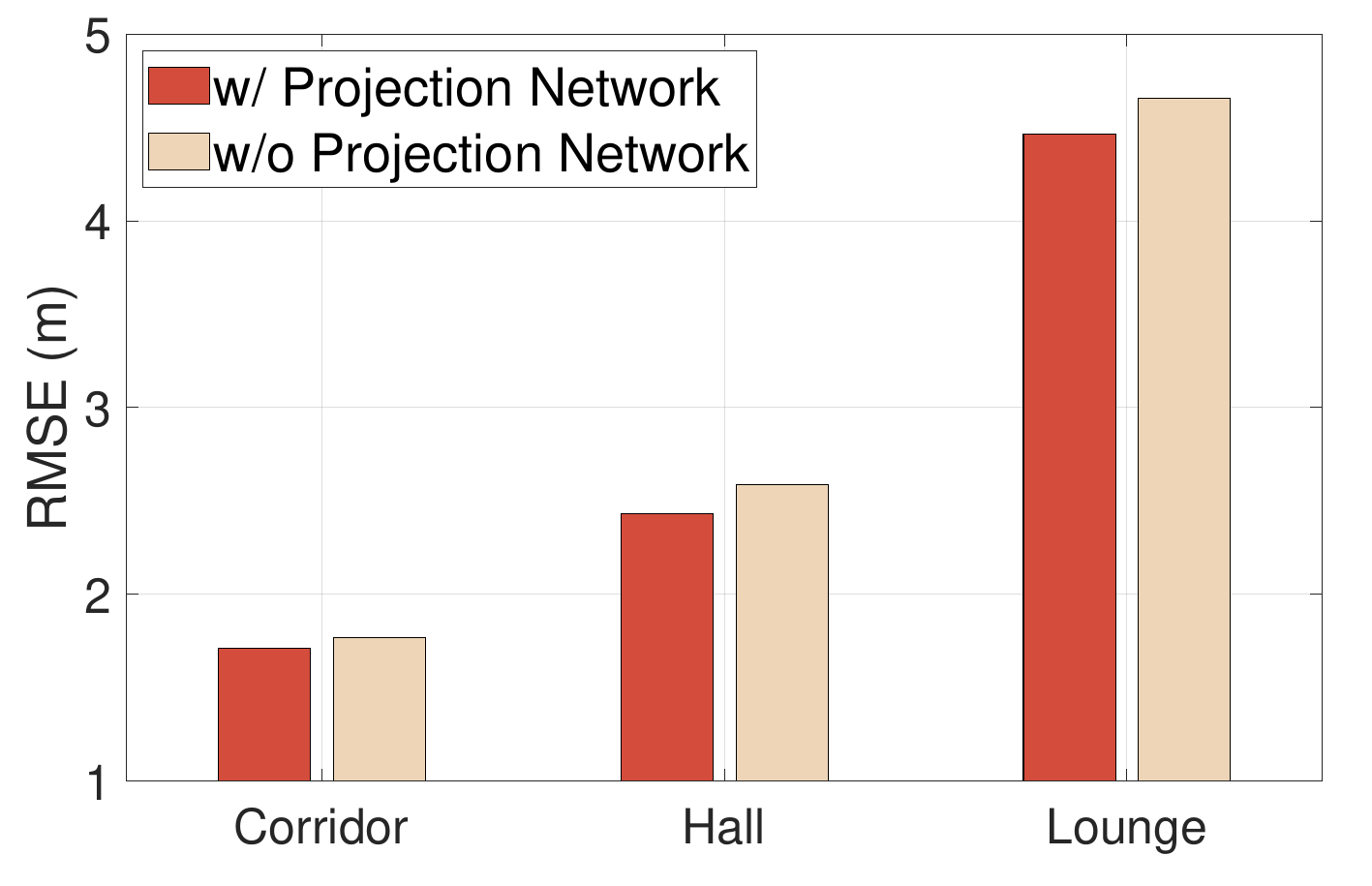}
		\caption{Localization Accuracy with/without the feature projection.}
		\label{projection}
	\end{minipage}
\end{figure*}

\subsection{Effect and Effectiveness  Analysis}
We further evaluate the effect and effectiveness of key modules in the proposed CSSLoc scheme.

\subsubsection{Effect on Limited CSI Fingerprints}
To evaluate the effect of localization performance with limited CSI fingerprints, we randomly choose the different numbers of RP to collect CSI fingerprints with the corresponding location information, and train the linear location predictor for location estimations in the specific indoor scenario. The deployed density is defined as $\rho=\frac{N_{lim}}{N_{all}}$, where $N_{lim}$ is the number of the deployed RP and $N_{all}$ is the total number of RP in the testbed indoor scenario. We compute the RMSE of location estimations in different indoor scenarios with a range of the deployed density as $\{0.2, 0.4, 0.6, 0.8, 1\}$. As illustrated in Fig. \ref{deploy-dense},  the localization errors seem to reduce with the increasing $\rho$, while the collection of CSI fingerprints from numerous RP is labor-intensive with cumbersome deployment efforts. From the experimental results, the best balance between the deployment efforts and the localization accuracy exists with the deployed density as $\rho=0.6$.  The reasonable accuracy is attained with a half of site survey, and thus, the proposed CSSLoc can effectively reduce the deployment efforts with high localization performance.  

\subsubsection{Effect on CSI Image Numbers}

We evaluate the effect of  CSI image numbers per RP on learning the effective representation with the contrastive self-supervised pre-training model. The number of CSI images per RP varies from 10 to 60, and then the localization results from specific indoor scenarios are shown in Fig. \ref{ImageSample}. With the complicated wireless environments, CSSLoc has achieved better accuracy with only 10 CSI radio images per RP, which is an optimal tradeoff between the localization accuracy and the computational cost. The increasing number of CSI  radio images demands a greater amount of GPU memory and brings CSI  redundancy with environmental noises and unpredictable interferences. 
From our experimental results, CSSLoc can use a few CSI images to learn effective representations for the downstream tasks of location estimations. The default value is optimal with 10 CSI images per RP in our case.

\subsubsection{Effectiveness on Feature Projection}
To verify the effectiveness of the nonlinear transformation with the feature projection, we calculate the evaluated errors of downstream localization tasks in three typical indoor scenarios with and without (w/o) the feature projection. The experimental results are shown in Fig. \ref{projection}. With the feature projection on the proposed CSSLoc system, the localization accuracy is improved by 3.3\% in the Corridor, 6.3\% in the Hall, and 4.3\% in the Lounge, respectively. The learnable non-linear transformation is useful for our self-supervised pre-training model to improve the high-quality of the learned feature representations and then gain accuracy improvement in complicated indoor scenarios.

\subsection{Parameter Analysis on Contrastive Self-Supervised Pre-Training Model}
We also analyze the effects of the proposed CSSLoc system with contrastive self-supervised learning as follows.

\subsubsection{Training Batch Size}
We evaluate the localization errors with the different batch sizes of CSSLoc's pre-training in diverse indoor scenarios, as shown in Fig. \ref{para-ana}(a). When the batch size is 128 or 256, the gradient instability of the developed pre-training model results in model collapse \cite{MoCo}. With a batch size of 1024, the RMSE reaches its lowest errors in all indoor scenarios. However, the larger batch sizes (2048 and 4096) demonstrate the degradation of localization performance due to gradient randomness \cite{MoCov2}.  From these results, the batch size of 1024 with our self-supervised pre-training model is set as the default.

\subsubsection{Momentum Update}
With the different values of the momentum coefficients,  
Fig. \ref{para-ana}(b) shows the localization errors of the proposed CSSLoc system in different indoor scenarios. It performs well when the momentum coefficient is between 0.99 and 1. However, the training loss fails to converge, even with model collapse, when the value of the momentum coefficient is too small. The results indicate that the higher momentum value benefits the training of our self-supervised pre-training model for keeping the consistent similarity metric on learning feature representations in a twin-branch network structure. We set the default of the momentum coefficient as 0.99.

\subsubsection{Temperature Coefficient}
\begin{figure*}[tbp]
		\centering
        \subfigure[]{
		\includegraphics[width=0.32\textwidth]{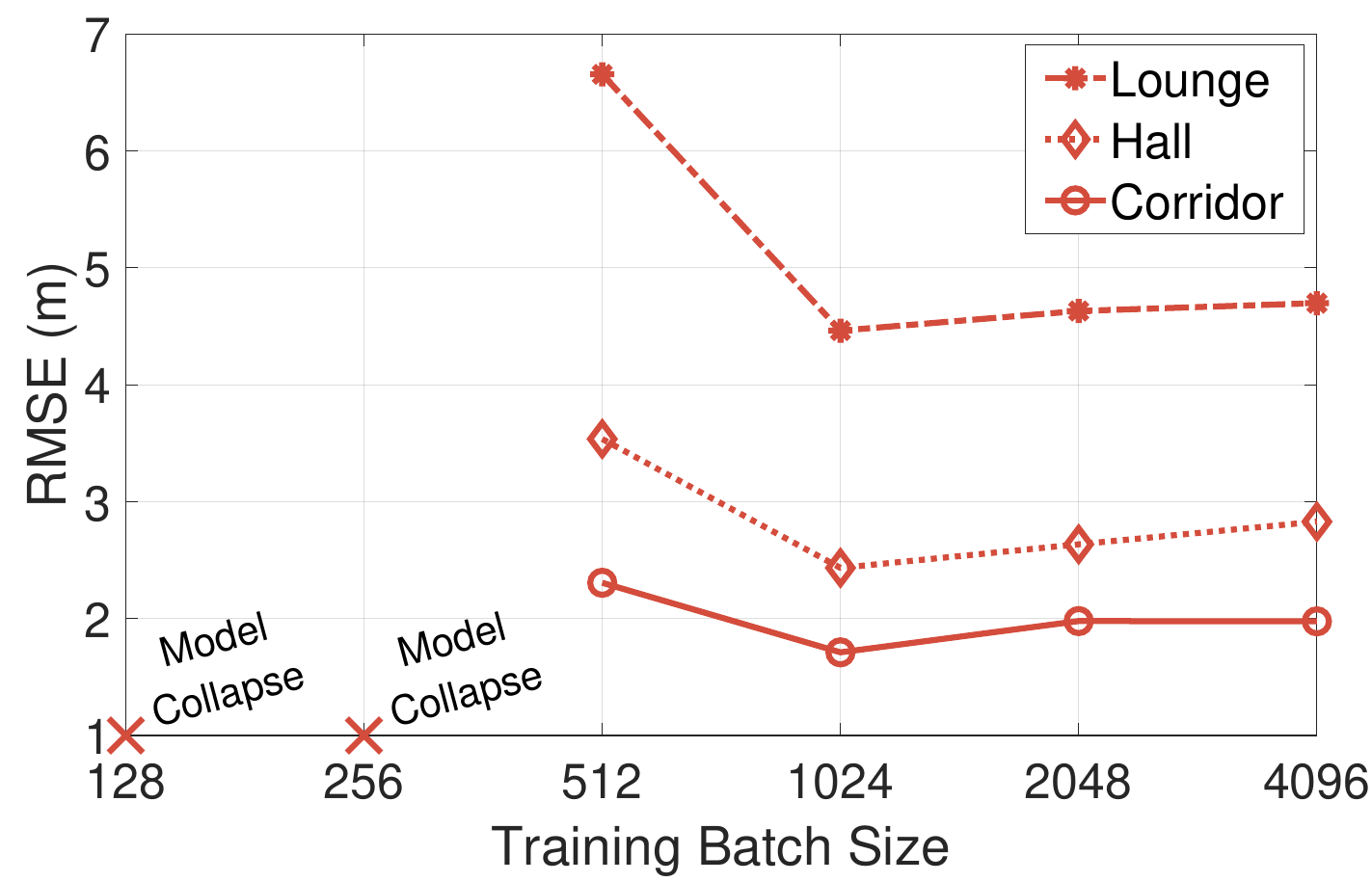}}
		\subfigure[]{
		\includegraphics[width=0.32\textwidth]{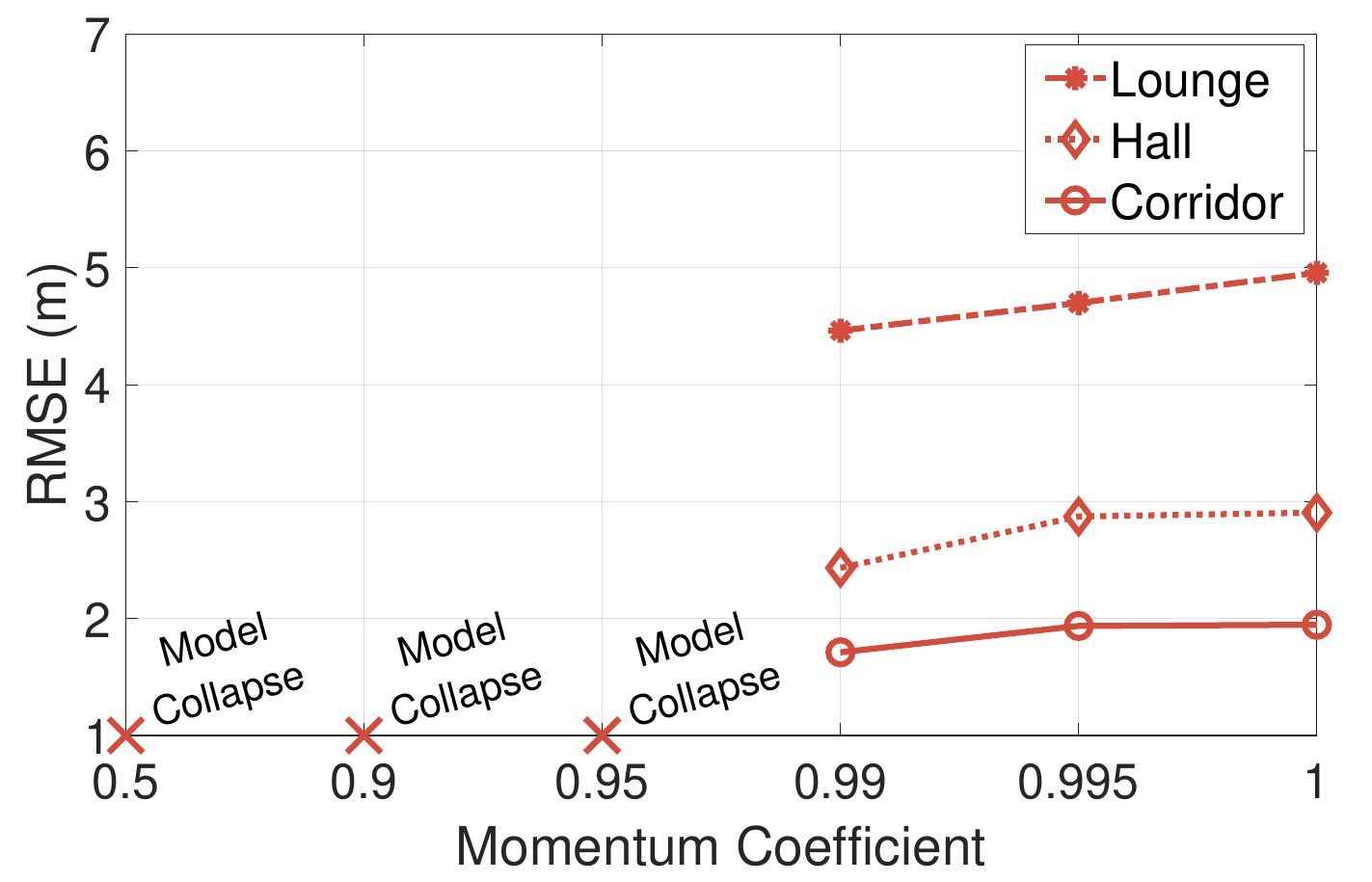}}
            \subfigure[]{
		\includegraphics[width=0.32\textwidth]{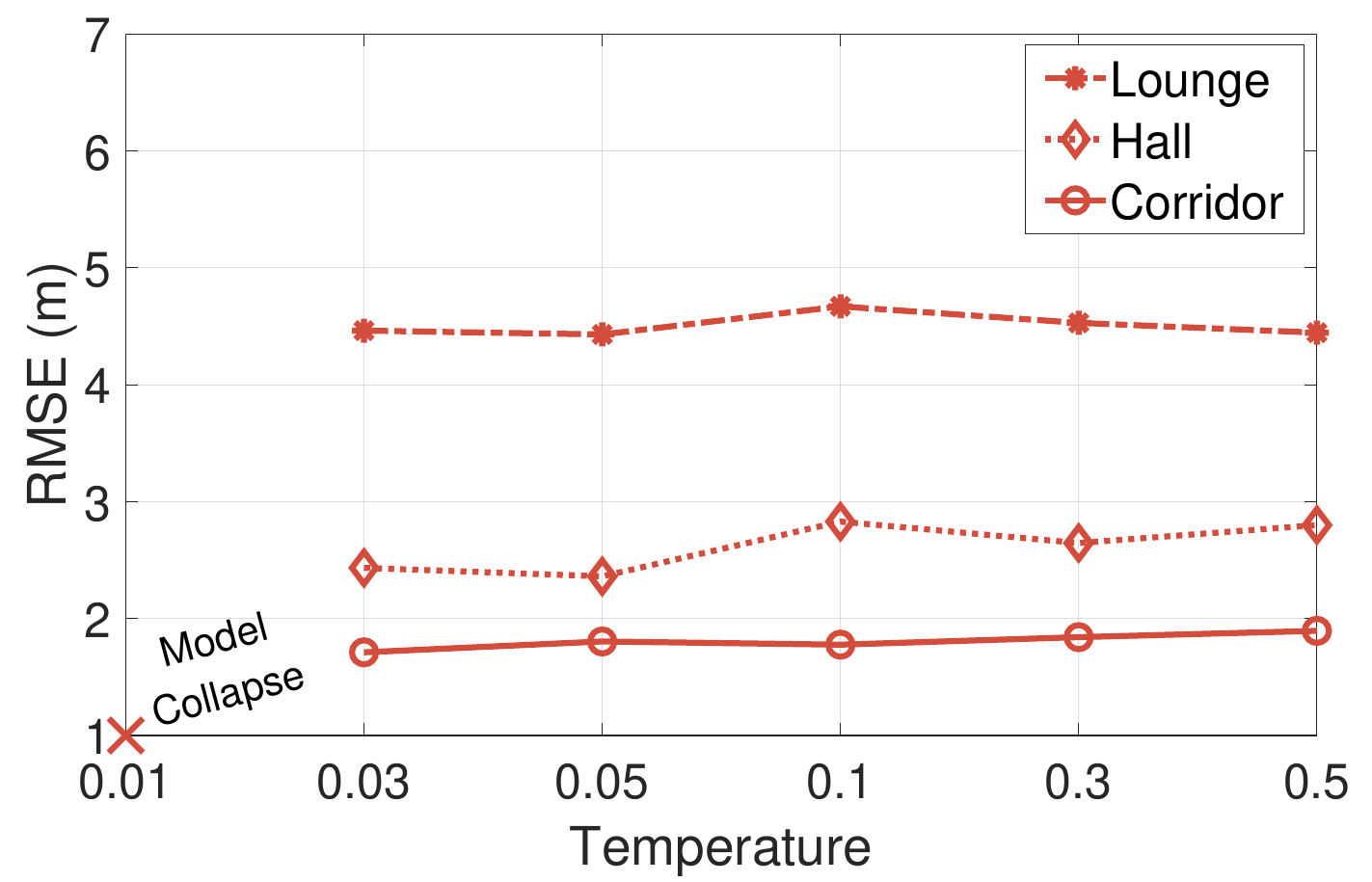}}
        \caption{Parameters Analysis of the Self-Supervised Pre-training Model.}
        \label{para-ana}
\end{figure*}

We examine a series of temperature coefficients to evaluate the effect of localization accuracy in different indoor scenarios. As shown in Fig. \ref{para-ana}(c), the model collapses are caused when the temperature is too low (e.g., 0.01). However, the increasing values of the temperature parameters tend to the higher errors. To combat this dilemma, CSSLoc requires an appropriately tuned temperature to attain the effectiveness of feature representation learning. From our experimental results, the default temperature value is 0.03. 
\section{Conclusion}
In this paper, we propose a novel framework of deep-learning-based localization with contrastive self-supervised representation learning, as CSSLoc that can achieve robust cross-scenario localization with high performance. Without the location information supervision, 
CSSLoc attempts to learn the characteristics of the similarities and dissimilarities underlying the radio data in a scenario-agnostic way that the similar samples are closely clustered together and different samples are far away in the representation space. For downstream localization tasks, the learned representations are directly transferred, and the location predictor is trained by limited CSI fingerprints, effectively reducing the deployment efforts. With extensive experimental results in typical indoor scenarios, CSSLoc can achieve a median accuracy of 1.97m with the robustness of environmental dynamics, which outperforms classical and state-of-the-art schemes at lower deployment and maintenance costs.

\end{document}